\catcode`\@=11  
\def\jnl{f} 
\def\revthree{t }    
\def\revfour{f }     
\def\iop{i }         
\def\nojnl{n }       

\if\jnl\iop 
 \documentclass[12pt]{iopart} 
 \usepackage{epsf}  
\fi 

\if\jnl\revfour 
 \documentclass[aps,draft,twocolumn,eqsecnum]{revtex4} 
 \usepackage{epsf}  
\fi 

\if\jnl\revthree 
\documentstyle[aps,eqsecnum,prd,epsf]{revtex} 
\fi 

\if\jnl\nojnl 
\documentstyle[12pt]{article} 
\fi 

\begin{document} 

\title{ 
Advantages of modified ADM formulation:\\ 
constraint propagation analysis of 
Baumgarte-Shapiro-Shibata-Nakamura system} 
\if\jnl\revfour 
\author{Gen Yoneda}\email{yoneda@mse.waseda.ac.jp} 
\author{Hisa-aki Shinkai}\email{hshinkai@postman.riken.go.jp} 
\affiliation{ 
${}^\ast$ Department of Mathematical Sciences, Waseda University, 
Okubo, Shinjuku, Tokyo,  169-8555, Japan 
\\ 
${}^\dagger$ Computational Science Division 
Institute of Physical \& Chemical Research (RIKEN), \\ 
Hirosawa, Wako, Saitama, 351-0198 Japan 
} 
\fi 

\if\jnl\revthree 
\author{Gen Yoneda ${}^\dagger$ and Hisa-aki Shinkai ${}^\ddagger$} 
\address{ 
{\tt yoneda@mse.waseda.ac.jp} ~~~ {\tt hshinkai@postman.riken.go.jp} 
\\ 
${}^\dagger$ Department of Mathematical Sciences, Waseda University, 
Okubo, Shinjuku, Tokyo,  169-8555, Japan 
\\ 
${}^\dagger$ Computational Science Division, 
Institute of Physical \& Chemical Research (RIKEN), \\ 
Hirosawa, Wako, Saitama, 351-0198 Japan 
} 
\fi 

 \date{September 29, 2002 (gr-qc/0204002) ~~ to appear in Phys. Rev. D.} 
\begin{abstract} 
Several numerical relativity groups are using a modified 
ADM formulation 
for their simulations, which was 
developed by Nakamura et al
(and widely cited as Baumgarte-Shapiro-Shibata-Nakamura system). 
This so-called BSSN formulation is shown to be more stable 
than the standard ADM 
formulation in many cases, 
and there have been many attempts to explain why 
this re-formulation has such an advantage. 
We try to explain the background mechanism of the BSSN equations 
by using eigenvalue analysis 
of constraint propagation equations. 
This analysis has been applied and has succeeded in explaining 
other systems in our series of works.
We derive the full set of the constraint propagation equations, 
and study it in the flat background space-time. 
We carefully examine how the replacements 
and adjustments in the equations change the propagation 
structure of the constraints, i.e. 
whether violation of constraints (if it exists) will decay or propagate away. 
We conclude that the better stability of
the BSSN system is obtained by their adjustments in the equations, 
and that the combination of the adjustments is in a good balance, i.e.  
a lack of their adjustments might fail to obtain the present stability. 
We further propose other adjustments to  
the equations,  which may offer more stable features than the current BSSN
equations.  
\end{abstract} 

\maketitle 

\section{Introduction} 

One of the current most important topics in the field  of 
numerical relativity 
is to find a formulation of the Einstein equations which gives 
us stable and accurate longterm  evolution. 
We all know that 
simulating space-time and matter based on general relativity is 
the essential research direction to go in the future, but we do not 
have a definite recipe for controlling  numerical blow-ups. 
We concentrate our discussion on free evolution of the Einstein 
equations based on the 3+1 (space + time) 
decomposition of  space-time, 
which requires solving the constraints only on the initial hypersurface 
and monitors the violation (error) of the calculation 
by checking constraints during the evolution. 

Over the decades, the Arnowitt-Deser-Misner (ADM) \cite{adm} formulation 
has been treated as 
the default for numerical relativists.  (More precisely, the version 
introduced by Smarr and York \cite{ADM-York} 
was taken as the default, 
which we denote the standard ADM formulation hereafter.) 
Although the ADM formulation mostly works for gravitational collapses or 
cosmological models 
in numerical treatments, it does not satisfy the requirement 
for longterm 
evolution e.g. the studies of gravitational wave sources. 

As we mentioned in our previous paper \cite{adjADMsch}, 
we think we can classify the 
current efforts of formulating equations for numerical relativity 
in the following three ways: 
(1) apply a modified ADM (BSSN) formulation \cite{SN,BS}, 
(2) apply a first-order hyperbolic formulation 
(see the references e.g. in 
\cite{ronbun1,reviewLehner,KST}), 
or  (3) apply an asymptotically constrained system 
\cite{BFHR,SiebelHuebner,SY-asympAsh,ronbun2}. 

The first refers to using a modified ADM formulation, 
originally proposed by Nakamura in late 80s, and 
subsequently modified by Nakamura-Oohara and 
Shibata-Nakamura \cite{SN}. 
This introduces conformal decomposition of the ADM variables, 
a new variable for 
calculating Ricci curvature, and adjusts the equations of motion using 
constraints. 
The advantage of this formulation was re-introduced by 
Baumgarte and Shapiro \cite{BS}, and therefore this is often 
cited as the BSSN formulation, which we follow also. 
The BSSN equations are now widely used in the large scale numerical 
computations, 
including coalescence of binary neutron stars \cite{binaryNS} 
and binary black holes 
\cite{binaryBH}. 

The second and third efforts use similar modifications such as 
introductions of new variables and/or adjustments of the equations, 
but differ in their purposes: to construct a hyperbolic 
formulation or to construct a formulation 
which constraints will decay or propagate away. 
The latter is intended 
to control numerical evolution 
so as the constrained manifold is its attractor. 
While the hyperbolic formulations have been extensively studied 
in this direction, 
we think the worrisome point in the discussion is the 
treatment of the non-principal part which is ignored in the hyperbolic 
formulation. 
As Kidder, Scheel and Teukolsky \cite{KST} reported 
recently, 
unless we reduce the effect of the non-principal part of the equations 
we may not get an advantages of hyperbolic formulation for numerical 
results \cite{ronbun1,HernPHD}. 

Through the series of studies \cite{adjADMsch,ronbun1,ronbun2,adjADM}, 
we propose a systematic treatment for constructing a robust evolution 
system against perturbative error. 
We call it an asymptotically constrained (or asymptotically stable)
system if the error decays itself.  The idea is to adjust evolution 
equations using constraints  (we term it an adjusted system), and to decide 
the coefficients (multipliers) by analyzing constraint propagation equations. 
We propose to apply an eigenvalue analysis of 
the propagation equations of the constraints, especially  
in its Fourier components so as to include the non-principal part in
the analysis. 
The characters of eigenvalues will be changed according 
to the adjustments to the 
original evolution equations.
We conjectured 
that the 
constraint violation which was occurred during the evolution will decay 
(if negative real eigenvalues) 
or propagate away (if pure imaginary eigenvalues). 

This conjecture was affirmatively confirmed to explain 
the numerical behaviors: 
wave propagation in the Maxwell equations \cite{ronbun2}, 
in the Ashtekar version of 
the Einstein equations \cite{ronbun2}, and in the ADM 
formulation (flat space-time background) \cite{adjADM}. 
The advantage of this construction scheme is that it can be applied to 
a formulation which is not a first-order hyperbolic form, 
such as to the ADM formulation \cite{adjADM,adjADMsch}. 
We think therefore our proposal 
is an alternative way to  
control/predict the 
violation of constraints. 
(We believe that the idea of the constraint propagation 
analysis first appeared in Frittelli \cite{Fri-con}, where she made 
hyperbolicity classification 
for the standard ADM formulation).

The purpose of this article is to apply this constraint propagation 
analysis to the BSSN formulation, 
and understand how each improvement contributes 
to more stable numerical evolution. 
Together with numerical comparisons with the 
standard ADM case \cite{potsdam0003,LHG}, 
this topic has been studied by many groups with different approaches. 
Using numerical test evolutios, 
Alcubierre et al. \cite{potsdam9908} found that 
the essential improvement is in 
the process of replacing terms by constraints, and that 
the eigenvalues of the BSSN {\it evolution equations} has fewer 
``zero eigenvalues" 
than those of ADM, and they conjectured that the instability 
can be caused by 
``zero eigenvalues" that violate ``gauge mode". 
Miller \cite{Miller} applied von Neumann's stability analysis 
to the plane wave propagation, and reported that BSSN has a 
wider range of 
parameters that give us stable evolution. 
These studies provide some supports regarding the advantage of BSSN, 
while it was also shown 
an example of an ill-posed solution in BSSN (as well in ADM) 
\cite{FrittelliGomez}. 
(Inspired by BSSN's conformal decomposition, 
several related hyperbolic formulations have 
also been proposed \cite{ArBona,FR99,ABMS}.) 

We think our analysis will offer a new vantage point on the topic, and 
contribute an alternative understanding of its background. 
Consequently, we propose 
more effective improvement of the BSSN system which has not yet been tried
in  numerical simulations. 

The construction of this paper is as follows. We review the BSSN 
system in 
\S \ref{sec2}, and there also we discuss where the adjustments are 
applied.  In \S \ref{sec3}  we apply our constraint propagation analysis to 
show how each improvement works in the BSSN equations, and in \S \ref{sec4} we 
extend our study to seek a better formulation which might 
be obtained by small steps.  We only consider the vacuum space-time 
thoughout 
the article, but the inclusion of matter is straightforward. 

\section{BSSN equations and their constraint propagation equations}
\label{sec2}
\subsection{BSSN equations} 

We start presenting the standard ADM formulation, 
which expresses the space-time 
with a pair of 3-metric $\gamma_{ij}$ and extrinsic curvature $K_{ij}$. 
The evolution equations become 
\begin{eqnarray} 
\partial_t^A \gamma_{ij}&=& 
-2\alpha K_{ij}+D_i\beta_j+D_j\beta_i, 
\\ 
\partial_t^A K_{ij} &=& 
\alpha R^{ADM}_{ij}+\alpha K K_{ij}-2\alpha K_{ik}{K^k}_j 
-D_iD_j \alpha 
\nonumber \\&& 
+(D_i \beta^k) K_{kj} +(D_j \beta^k) K_{ki} 
+\beta^k D_k K_{ij} 
\end{eqnarray} 
where $\alpha, \beta_i$ are the lapse and shift function and $D_i$ 
is the covariant derivative on 3-space. 
The symbol $\partial_t^A$ means the time derivative defined 
by these equations, and 
we distinguish them from those of the BSSN equations $\partial_t^B$, 
which will 
be defined in (\ref{BSSNeqmPHI})-(\ref{BSSNeqmTG}). 
The associated constraints are 
the Hamiltonian constraint ${\cal H}$ and the momentum constraints 
${\cal M}_i$: 
\begin{eqnarray} 
{\cal H}^{ADM}&=& R^{ADM}+K^2-K_{ij}K^{ij}, 
\\ 
{\cal M}^{ADM}_i&=& 
D_j K^j{}_i-D_i K. 
\end{eqnarray} 

The widely used notation\cite{SN,BS} is to introduce the variables 
($\varphi,\tilde{\gamma}_{ij}$,$K$,$\tilde{A}_{ij}$,$\tilde{\Gamma}^i$) 
instead of ($\gamma_{ij}$,$K_{ij}$), where 
\begin{eqnarray} 
\varphi &=& (1/12)\log ({\rm det}\gamma_{ij}), 
\label{BSSNval1} 
\\ 
\tilde{\gamma}_{ij} & = & e^{-4\varphi}\gamma_{ij}, 
\label{BSSNval2} 
\\ 
K  &=& \gamma^{ij}K_{ij}, \label{BSSNval3} 
\\ 
\tilde{A}_{ij} &=& 
e^{-4\varphi}(K_{ij} - (1/3)\gamma_{ij}K), \label{BSSNval4} 
\\ 
\tilde{\Gamma}^i &=& 
\tilde{\Gamma}^i_{jk}\tilde{\gamma}^{jk}.
\label{BSSNval5} 
\end{eqnarray} 
The new variable $\tilde{\Gamma}^i$ was introduced in order to 
calculate Ricci curvature more accurately.  $\tilde{\Gamma}^i$ also 
contributes to make the system re-produce wave equations in its 
linear limit. 
In the BSSN formulation, Ricci curvature is not calculated as 
\begin{equation} 
R^{ADM}_{ij} 
= 
\partial_k \Gamma^k_{ij} 
-\partial_i\Gamma^k_{kj} 
+\Gamma^l_{ij}\Gamma^k_{lk} 
-\Gamma^l_{kj}\Gamma^k_{li}, 
\end{equation} 
but 
\begin{eqnarray} 
R^{BSSN}_{ij} 
&=& 
R^\varphi_{ij}+\tilde R_{ij}, 
\label{BSricci} 
\\ 
R^\varphi_{ij} 
&=& 
-2\tilde{D}_i\tilde{D}_j\varphi 
-2\tilde{\gamma}_{ij}\tilde{D}^k\tilde{D}_k\varphi 
\nonumber \\&& 
+4(\tilde{D}_i\varphi)(\tilde{D}_j\varphi) 
-4\tilde{\gamma}_{ij}(\tilde{D}^k\varphi)(\tilde{D}_k\varphi), 
\\ 
\tilde{R}_{ij} 
&=& 
-(1/2)\tilde{\gamma}^{lk}\partial_{l}\partial_{k}\tilde{\gamma}_{ij} 
+\tilde{\gamma}_{k(i}\partial_{j)}\tilde{\Gamma}^k 
+\tilde{\Gamma}^k\tilde{\Gamma}_{(ij)k} 
\nonumber \\&& 
+2\tilde{\gamma}^{lm}\tilde{\Gamma}^k_{l(i}\tilde{\Gamma}_{j)km} 
+\tilde{\gamma}^{lm}\tilde{\Gamma}^k_{im}\tilde{\Gamma}_{klj}, 
\end{eqnarray} 
where $\tilde{D}_i$ is covariant derivative associated 
with $\tilde{\gamma}_{ij}$. 
These are weakly equivalent, but $R^{BSSN}_{ij}$ does have 
wave operator apparently in the flat background limit, so that 
we can expect more natural wave propagation behavior. 

Additionally, BSSN requires us to impose the conformal factor as 
\begin{equation} 
\tilde{\gamma}(:={\rm det} \tilde{\gamma}_{ij})=1, \label{scalingdef} 
\end{equation} 
during the evolution.  This is a kind of definition, but can also be 
thought of as a constraint.  We will return to this point shortly. 

BSSN's improvements  are not only the introductions of 
new variables, 
but also the replacement of terms in the evolution equations using 
the constraints. 
The purpose of this article is to understand and to identify which 
improvement works for the stability. 
Before doing that 
we first show the standard set of the BSSN evolution equations: 
\begin{widetext} 
\begin{eqnarray} 
\partial_t^B \varphi 
&=& 
-(1/6)\alpha K+(1/6)\beta^i(\partial_i\varphi)+(\partial_i\beta^i), 
\label{BSSNeqmPHI} 
\\ 
\partial_t^B \tilde{\gamma}_{ij} 
&=& 
-2\alpha\tilde{A}_{ij} 
+\tilde{\gamma}_{ik}(\partial_j\beta^k) 
+\tilde{\gamma}_{jk}(\partial_i\beta^k) 
-(2/3)\tilde{\gamma}_{ij}(\partial_k\beta^k) 
+\beta^k(\partial_k\tilde{\gamma}_{ij}), 
\label{BSSNeqmtgamma} 
\\ 
\partial_t^B K 
&=& 
-D^iD_i\alpha 
+\alpha \tilde{A}_{ij}\tilde{A}^{ij} 
+(1/3) \alpha K^2 
+\beta^i (\partial_i K), 
\label{BSSNeqmK} 
\\ 
\partial_t^B \tilde{A}_{ij} 
&=& 
-e^{-4\varphi}(D_iD_j\alpha)^{TF} 
+e^{-4\varphi} \alpha (R^{BSSN}_{ij})^{TF} 
+\alpha K\tilde{A}_{ij} 
-2\alpha \tilde{A}_{ik}\tilde{A}^k{}_j 
+(\partial_i\beta^k)\tilde{A}_{kj} 
+(\partial_j\beta^k)\tilde{A}_{ki} 
\nonumber \\&& 
-(2/3)(\partial_k\beta^k)\tilde{A}_{ij} 
+\beta^k(\partial_k \tilde{A}_{ij}), 
\label{BSSNeqmTA} 
\\ 
\partial_t^B \tilde{\Gamma}^i &=& 
-2(\partial_j\alpha)\tilde{A}^{ij} 
+2\alpha 
\big(\tilde{\Gamma}^i_{jk}\tilde{A}^{kj} 
-(2/3)\tilde{\gamma}^{ij}(\partial_j K) 
+6\tilde{A}^{ij}(\partial_j\varphi) 
\big) 
-\partial_j 
\big( \beta^k(\partial_k\tilde{\gamma}^{ij}) 
-\tilde{\gamma}^{kj}(\partial_k\beta^{i}) 
\nonumber \\&& 
-\tilde{\gamma}^{ki}(\partial_k\beta^{j}) 
+(2/3)\tilde{\gamma}^{ij}(\partial_k\beta^k) 
\big). 
\label{BSSNeqmTG} 
\end{eqnarray} 
\end{widetext} 

We next summarize the constraints in this system. 
The normal Hamiltonian and momentum constraints 
(the ``kinematic" constraints) 
are naturally written as 
\begin{eqnarray} 
{\cal H}^{BSSN} 
&=& 
R^{BSSN}+K^2-K_{ij}K^{ij}, 
\label{BSSNconstraintH} 
\\ 
{\cal M}^{BSSN}_i 
&=& 
{\cal M}^{ADM}_i,  \label{BSSNconstraintM} 
\end{eqnarray} 
where we use Ricci scalar defined by (\ref{BSricci}). 
Additionally, we regard the following three as 
the constraints (the ``algebraic" constraints): 
\begin{eqnarray} 
{\cal G}^i &=& \tilde{\Gamma}^i-\tilde{\gamma}^{jk} 
\tilde{\Gamma}^i_{jk}, \label{BSSNconstraintG} 
\\ 
{\cal A}&=&\tilde{A}_{ij}\tilde{\gamma}^{ij}, \label{BSSNconstraintA} 
\\ 
{\cal S} &=& 
\tilde{\gamma}-1, 
\label{BSSNconstraintS} 
\end{eqnarray} 
where the first two are from the algebraic 
definition of the variables (\ref{BSSNval4}) and (\ref{BSSNval5}), 
and the (\ref{BSSNconstraintS}) is from the 
requirement of (\ref{scalingdef}). 
Hereafter we write ${\cal H}^{BSSN}$ and ${\cal M}^{BSSN}$ 
simply as ${\cal H}$ and ${\cal M}$ respectively. 

Taking careful account of these constraints, 
(\ref{BSSNconstraintH}) and (\ref{BSSNconstraintM}) can be expressed 
directly as 
\begin{widetext} 
\begin{eqnarray} 
{\cal H} &=& 
e^{-4\varphi}\tilde{R} 
-8e^{-4\varphi}\tilde{D}^j\tilde{D}_j\varphi 
-8e^{-4\varphi}(\tilde{D}^j\varphi)(\tilde{D}_j\varphi) 
+(2/3)K^2 
-\tilde{A}_{ij}\tilde{A}^{ij} 
-(2/3) {\cal A} K, 
\label{BSSNconstraint1} 
\\ 
{\cal M}_i 
&=& 
6\tilde{A}^j{}_{i}(\tilde{D}_j \varphi) 
-2{\cal A}(\tilde{D}_i \varphi) 
-(2/3) (\tilde{D}_i K) 
+\tilde{\gamma}^{kj}(\tilde{D}_j\tilde{A}_{ki}). 
\label{BSSNconstraint2} 
\end{eqnarray} 

In summary, the fundamental dynamical variables in BSSN are 
($\varphi,\tilde{\gamma}_{ij}$, 
$K$,$\tilde{A}_{ij}$,$\tilde{\Gamma}^i$), 
total 17. 
The gauge quantities are ($\alpha, \beta^i$) which is 4, and the 
constraints are 
$({\cal H},{\cal M}_i,{\cal G}^i,{\cal A},{\cal S})$, 
i.e. 9 components. 
As a result, 4 (2 by 2) 
components are left which correspond to 
two gravitational polarization modes. 

\subsection{Adjustments in evolution equations} 

Next, we show the BSSN evolution equations 
(\ref{BSSNeqmPHI})-(\ref{BSSNeqmTG})  again, 
identifying where the terms 
are replaced using the constraints, 
(\ref{BSSNconstraintH})-(\ref{BSSNconstraintS}). 

By a straightforward 
calculation,  we get: 
\begin{eqnarray} 
\partial_t^B \varphi&=& 
\partial_t^A \varphi 
+(1/6)\alpha{\cal A} 
-(1/12)\tilde{\gamma}^{-1}(\partial_j{\cal S})\beta^j, 
\label{BSSNeq1} 
\\ 
\partial_t^B \tilde{\gamma}_{ij}&=& 
\partial_t^A \tilde{\gamma}_{ij} 
-(2/3)\alpha \tilde{\gamma}_{ij}{\cal A} 
+(1/3)\tilde{\gamma}^{-1}(\partial_k{\cal S}) 
\beta^k\tilde{\gamma}_{ij}, 
\label{BSSNeq2} 
\\ 
\partial_t^B K 
&=& 
\partial_t^A K 
-(2/3)\alpha K {\cal A} 
-\alpha {\cal H} 
+\alpha e^{-4\varphi}(\tilde{D}_j{\cal G}^j), 
\label{BSSNeq3} 
\\ 
\partial_t^B \tilde{A}_{ij} 
&=& 
\partial_t^A \tilde{A}_{ij} 
+\big( 
(1/3)\alpha \tilde{\gamma}_{ij}K 
-(2/3)\alpha \tilde{A}_{ij}\big) 
{\cal A} 
+\big( 
(1/2)\alpha e^{-4\varphi}(\partial_k\tilde{\gamma}_{ij}) 
-(1/6)\alpha e^{-4\varphi} 
\tilde{\gamma}_{ij}\tilde{\gamma}^{-1}(\partial_k{\cal S}) 
\big) 
{\cal G}^k 
\nonumber \\&& 
+\alpha e^{-4\varphi}\tilde{\gamma}_{k(i}(\partial_{j)}{\cal G}^k) 
-(1/3)\alpha e^{-4\varphi}\tilde{\gamma}_{ij}(\partial_k{\cal G}^k), 
\label{BSSNeq4} 
\\ 
\partial_t^B\tilde{\Gamma}^i 
&=& 
\partial_t^A\tilde{\Gamma}^i 
+\big( 
-(2/3)(\partial_j \alpha) \tilde{\gamma}^{ji} 
-(2/3)\alpha (\partial_j \tilde{\gamma}^{ji}) 
-(1/3)\alpha \tilde{\gamma}^{ji}\tilde{\gamma}^{-1} 
(\partial_j{\cal S}) 
+4\alpha\tilde{\gamma}^{ij}(\partial_j \varphi) 
\big){\cal A} 
-(2/3)\alpha \tilde{\gamma}^{ji}(\partial_j {\cal A}) 
\nonumber \\&& 
+2\alpha\tilde{\gamma}^{ij}{\cal M}_j 
-(1/2)(\partial_k\beta^i)\tilde{\gamma}^{kj} 
\tilde{\gamma}^{-1}(\partial_j{\cal S}) 
+(1/6)(\partial_j\beta^k)\tilde{\gamma}^{ij} 
\tilde{\gamma}^{-1}(\partial_k{\cal S}) 
+(1/3)(\partial_k\beta^k)\tilde{\gamma}^{ij} 
\tilde{\gamma}^{-1}(\partial_j{\cal S}) 
\nonumber \\&& 
+(5/6)\beta^k\tilde{\gamma}^{-2}\tilde{\gamma}^{ij} 
(\partial_k{\cal S})(\partial_j{\cal S}) 
+(1/2)\beta^k\tilde{\gamma}^{-1}(\partial_k 
\tilde{\gamma}^{ij})(\partial_j{\cal S}) 
+(1/3)\beta^k\tilde{\gamma}^{-1}(\partial_j 
\tilde{\gamma}^{ji})(\partial_k{\cal S}). 
\label{BSSNeq5} 
\end{eqnarray} 
\end{widetext} 
where $\partial_t^A $ denotes the part of no replacements, i.e. 
the terms only use the standard ADM evolution equations in its 
time derivatives.

{}From (\ref{BSSNeq1})-(\ref{BSSNeq5}), we understand that 
all the BSSN evolution equations are {\it adjusted} using constraints. 
This fact will give us the importance of the scaling constraint 
${\cal S}=0$ 
and the tracefree operation ${\cal A}=0$ during the evolution. 

As we have pointed out in the case of adjusted ADM systems 
\cite{adjADM,adjADMsch}, 
certain combinations of adjustments (replacements) 
in the evolution equations 
change the eigenvalues of constraint propagation equations drastically.  
For example, all negative eigenvalues can be negative real by applying
Detweiler's adjustment \cite{detweiler} or its simplified version. 
One common fact we found is that such a case has an adjustment which 
breaks time reversal parity of the original equation. 
That is, with a change of time 
integration direction 
$\partial_t \rightarrow - \partial_t$, an adjusted term might become 
effective if it breaks time reversal symmetry. (This time asymmtric feature 
was first impremented as a ``lambda-system" in \cite{BFHR}.) 
Unfortunately, for the case of the BSSN equations, 
(\ref{BSSNeq1})-(\ref{BSSNeq5}), 
all the above adjustments keep the time reversal symmetry. 
So that we can not expect direct decays of constraint 
violation in the present form. 
We will give the details on this point later.

\section{Constraint propagation analysis in flat space-time} 
\label{sec3}
\subsection{Procedures} 
We start this section overviewing the procedures and our goals. 
In our series of previous works \cite{ronbun2,adjADM,adjADMsch}, we have 
concluded that eigenvalue analysis of the constraint propagation 
equations are quite useful 
for explaining or predicting how the constraint violation grows. 

Suppose we have a set of dynamical variables $u^a (x^i,t)$, and their 
evolution equations 
\begin{equation} 
\partial_t u^a = f(u^a, \partial_i u^a, \cdots), \label{ueq} 
\end{equation} 
and the 
(first class) constraints 
\begin{equation} 
C^\alpha (u^a, \partial_i u^a, \cdots) \approx 0. 
\end{equation} 
For monitoring the violation of constraints, we propose to investigate 
the evolution equations of $C^\alpha$ (constraint propagation), 
\begin{equation} 
\partial_t C^\alpha = g(C^\alpha, \partial_i C^\alpha, \cdots). 
\label{Ceq} 
\end{equation} 
(We do not mean to integrate (\ref{Ceq}) numerically, 
but rather to evaluate them analytically in advance.) 
In order to analyze the contributions of all RHS 
terms in (\ref{Ceq}), 
we propose to reduce (\ref{Ceq}) in ordinary differential equations 
by Fourier 
transformation, 
\begin{eqnarray} 
\partial_t \hat{C}^\alpha &=& \hat{g}(\hat{C}^\alpha) 
=M^\alpha{}_{\beta} \hat{C}^\beta, 
\label{CeqF} 
\end{eqnarray} 
where $C(x,t)^\rho 
={\int} \hat{C}(k,t)^\rho\exp(ik\cdot x)d^3k$, 
and then to analyze the set of eigenvalues, say $\Lambda_\alpha$,  of the 
coefficient matrix, $M^\alpha{}_{\beta}$, in 
(\ref{CeqF}). 
We call $\Lambda$s and $M^\alpha{}_{\beta}$
the constraint amplification factors (CAFs) of
(\ref{Ceq}) and constraint propagation matrix, respectively.  
Our guidelines to have
`better stability' are that 
\begin{itemize} 
\item[(A)] If the CAFs have a 
{\it negative real-part } 
(the constraints are forced to be 
diminished), then we see 
more stable evolution than a system 
which has a positive real-part. 
\item[(B)] If the CAFs have a 
{\it non-zero imaginary-part } 
(the constraints are propagating away), then we see 
more stable evolution than a system which has 
zero CAFs. 
\end{itemize} 
We found heuristically that the system becomes more stable when 
more $\Lambda$s satisfy the above criteria \cite{ronbun1,ronbun2}. 
We note that these guidelines are confirmed numerically for 
wave propagation in the Maxwell system and in the Ashtekar version of 
the Einstein system \cite{ronbun2}, and also for error propagation 
in Minkowskii space-time using adjusted ADM systems \cite{adjADM}. 
Supporting theorems for above (A) was recently discussed \cite{diagCP}.  


The above features of the constraint propagation, (\ref{Ceq}), 
will differ when we modify the original evolution equations. 
Suppose we add (adjust) the evolution equations using constraints 
\begin{equation} 
\partial_t u^a = f(u^a, \partial_i u^a, \cdots) 
+ F(C^\alpha, \partial_i C^\alpha, \cdots), \label{DeqADJ} 
\end{equation} 
then (\ref{Ceq}) will also be modified as 
\begin{equation} 
\partial_t C^\alpha = g(C^\alpha, \partial_i C^\alpha, \cdots) 
+ G(C^\alpha, \partial_i C^\alpha, \cdots). \label{CeqADJ} 
\end{equation} 
Therefore, the problem is how to adjust the evolution equations 
so that their constraint propagation satisfies the above criteria 
as much as possible.

\subsection{BSSN constraint propagation equations } 

Our purpose in this section is to apply the above procedure to 
the BSSN system. 
The set of the constraint propagation equations, 
$\partial_t ({\cal H}, {\cal M}_i, 
{\cal G}^i,  {\cal A}, {\cal S})^T$, 
turns to be quite long and not elegant 
(is not a first-order hyperbolic 
and includes many non-linear terms), and we put them in Appendix. 
In order to understand the fundamental structure, we hereby show an 
analysis on the flat space-time background. 

For the flat background metric $g_{\mu\nu}=\eta_{\mu\nu}$, 
the first order perturbation equations of 
(\ref{BSSNeq1})-(\ref{BSSNeq5})  can be written as 
\begin{widetext} 
\begin{eqnarray} 
\partial_t{}^{\!(1)\!\!}\phi &=& 
-(1/6){}^{\!(1)\!\!} K 
+(1/6)(\kappa_{\phi}-1){}^{\!(1)\!\!}{\cal A}, 
\label{eqmMinkow1} 
\\ 
\partial_t{}^{\!(1)\!\!}\tilde \gamma_{ij}&=& 
-2{}^{\!(1)\!\!}\tilde A_{ij} 
-(2/3)(\kappa_{\tilde \gamma}-1)\delta_{ij}{}^{\!(1)\!\!}{\cal A}, 
\label{eqmMinkow2} 
\\ 
\partial_t {}^{\!(1)\!\!} K 
&=& 
-(\partial_j \partial_j {}^{\!(1)\!\!}\alpha) 
+(\kappa_{K1}-1)\partial_j{}^{\!(1)\!\!}{\cal G}^j 
-(\kappa_{K2}-1){}^{\!(1)\!\!}{\cal H}, 
\label{eqmMinkow3} 
\\ 
\partial_t {}^{\!(1)\!\!} \tilde A_{ij} 
&=& 
{}^{\!(1)\!\!}(R^{BSSN}_{ij}){}^{TF} 
-{}^{\!(1)\!\!}(\tilde D_i \tilde D_j \alpha){}^{TF} 
+(\kappa_{A1}-1)\delta_{k(i}(\partial_{j)}{}^{\!(1)\!\!}{\cal G}^k) 
-(1/3)(\kappa_{A2}-1)\delta_{ij}(\partial_k {}^{\!(1)\!\!} {\cal G}^k), 
\label{eqmMinkow4} 
\\ 
\partial_t {}^{\!(1)\!\!} \tilde\Gamma^i &=& 
-(4/3)(\partial_i {}^{\!(1)\!\!} K) 
-(2/3)(\kappa_{\tilde \Gamma 1}-1)(\partial_i{}^{\!(1)\!\!}{\cal A}) 
+2(\kappa_{\tilde\Gamma 2}-1){}^{\!(1)\!\!}{\cal M}_i, 
\label{eqmMinkow5} 
\end{eqnarray} 
where we introduced parameters $\kappa$s, 
all $\kappa=0$ reproduce no adjustment 
case from the standard ADM equations, 
and  all $\kappa=1$ correspond to the BSSN 
equations. 
We express them as 
\begin{eqnarray} 
\kappa_{adj}&:=&(\kappa_{\varphi},\kappa_{\tilde{\gamma}}, 
\kappa_{K1}, 
\kappa_{K2}, 
\kappa_{A1}, 
\kappa_{A2}, 
\kappa_{\tilde{\Gamma} 1}, 
\kappa_{\tilde{\Gamma} 2}). 
\end{eqnarray}

Constraint propagation equations at the first order 
in the flat space-time, then, become: 
\begin{eqnarray} 
\partial_t{}^{\!(1)\!\!}{\cal H}&=& 
\left(\kappa_{\tilde{\gamma}}-(2/3)\kappa_{\tilde{\Gamma} 
1}-(4/3)\kappa_{\varphi}+2\right) 
\partial_j\partial_j{}^{\!(1)\!\!}{\cal A} 
+2(\kappa_{\tilde{\Gamma} 2}-1)(\partial_j{}^{\!(1)\!\!}{\cal M}_j), 
\label{CPMinkow1} 
\\ 
\partial_t {}^{\!(1)\!\!} {\cal M}_i&=& 
\left 
(-(2/3)\kappa_{K1}+(1/2)\kappa_{A1}-(1/3)\kappa_{A2}+(1/2)\right) 
\partial_i\partial_j{}^{\!(1)\!\!}{\cal G}^j 
\nonumber \\ && 
+(1/2)\kappa_{A1}\partial_j\partial_j{}^{\!(1)\!\!}{\cal G}^i 
+\left((2/3)\kappa_{K2}-(1/2)\right) 
\partial_i{}^{\!(1)\!\!}{\cal H}, 
\\ 
\partial_t {}^{\!(1)\!\!} {\cal G}^i 
&=& 
2\kappa_{\tilde{\Gamma} 2}{}^{\!(1)\!\!}{\cal M}_i 
+(-(2/3)\kappa_{\tilde{\Gamma} 
1}-(1/3)\kappa_{\tilde{\gamma}})(\partial_i{}^{\!(1)\!\!}{\cal A}), 
\\ 
\partial_t {}^{\!(1)\!\!} {\cal S}&=& 
-2\kappa_{\tilde{\gamma}}{}^{\!(1)\!\!}{\cal A}, 
\\ 
\partial_t {}^{\!(1)\!\!} {\cal A}&=& 
(\kappa_{A1}-\kappa_{A2})(\partial_{j}{}^{\!(1)\!\!}{\cal G}^j). 
\label{CPMinkow5} 
\end{eqnarray} 

We will discuss CAFs  of 
(\ref{CPMinkow1})-(\ref{CPMinkow5}). 

\end{widetext}


\subsection{Effect of adjustments}\label{sec3c} 
We check CAFs of the BSSN equations in detail.  The list of examples 
is shown also in Table \ref{table1}.  Hereafter we let 
$k^2=k_x^2+k_y^2+k_z^2$ for Fourier wave numbers. 
\begin{enumerate} 
\item The no-adjustment case, $\kappa_{adj}=$(all zeros). This is the 
starting point of the discussion.  In this case, 
$$ 
CAFs=(0 \,  (\times 7),\pm \sqrt{-k^2}), 
$$ 
i.e., $(0 \, (\times 7), \pm \mbox{pure imaginary} 
\, \mbox{(1 pair)})$. 
In the standard ADM formulation, which uses $(\gamma_{ij}, K_{ij})$, 
CAFs are $(0, 0, \pm \mbox{Pure Imaginary})$ \cite{adjADM}. 
Therefore if we do not apply adjustments in the BSSN equations the
constraint propagation structure is quite similar to that of 
the standard ADM. 
\item For the BSSN equations,  $\kappa_{adj}=$(all 1s), 
$$ 
CAFs=(0 \,  (\times 3),\pm \sqrt{-k^2} \, \mbox{(3 pairs)}), 
$$ 
i.e., $(0 \, (\times 3), \pm \mbox{Pure Imaginary} 
\, \mbox{(3 pairs)})$. 
The number of pure imaginary CAFs is increased over that of No.1, 
and we 
conclude this is the advantage of adjustments used in the BSSN equations. 
\item No ${\cal S}$-adjustment case. 
All the numerical experiments so far 
apply the scaling condition ${\cal S}$ 
for the conformal factor $\varphi$. 
The ${\cal S}$-originated terms appear many places in the BSSN equations 
(\ref{BSSNeqmPHI})-(\ref{BSSNeqmTG}), so that we suspect non-zero
${\cal S}$ is a kind of source of the constraint violation. 
However,
since all ${\cal S}$-originated terms do not appear in the 
flat space-time background analysis, 
[no adjusted terms in (\ref{eqmMinkow1})-(\ref{eqmMinkow5})],
our analysis is independent of the ${\cal S}$-constraint. 
(Remark that we do not deny the effect of ${\cal S}$-adjustment
 in other situation.) 
\item No ${\cal A}$-adjustment case. 
The trace (or traceout) condition for 
the variables is also considered necessary (e.g. \cite{AB2001}). 
This can be checked with 
$\kappa_{adj}=(\kappa,\kappa, 1,1, 1,1, \kappa,1)$, and we get 
$$ 
CAFs=( 0 \, (\times 3), \pm \sqrt{-  k^2} \, \mbox{(3 pairs)} ), 
$$ 
independent of $\kappa$. 
Therefore the effect of ${\cal A}$-adjustment 
is unimportant according to this analysis, i.e. 
on flat space-time background.
(Remark that we do not deny the effect of ${\cal A}$-adjustment
 in other situation.)
\item No ${\cal G}^i$-adjustment case. 
The introduction of $\Gamma^i$ is the 
key in the BSSN system.  
If we do not apply adjustments by ${\cal G}^i$,
($\kappa_{adj}=(1,1,0,1,0,0,1,1)$) then we get
$$ 
CAFs=( 0 (\times 7), \pm 
\sqrt{  - k^2  } ), 
$$ 
which is the same with No.1. 
That is, adjustments due to ${\cal G}^i$ terms 
are effective to make a progress from ADM. 
\item No ${\cal M}_i$-adjustment case.  This can be checked with 
$\kappa_{adj}=( 1,1,1,1,1,1,1,\kappa )$, and we get 
\begin{eqnarray*} 
CAFs&=&( 0, \pm \sqrt{-  \kappa k^2} \, \mbox{(2 pairs)}, 
\nonumber \\&& 
\pm \sqrt{ - k^2 ( -1 + 4 \kappa + |1-4\kappa|)/6}, 
\nonumber \\&& 
\pm \sqrt{ - k^2 ( -1 + 4 \kappa -  |1-4\kappa|)/6} 
). 
\end{eqnarray*} 
If $\kappa=0$, then 
$(0 (\times 7), \pm 
\sqrt{  k^2 /3 } )$, which is $( 0 (\times 7), \pm 
\mbox{real value} )$.  Interestingly, 
these real values indicate the existence of 
the error growing mode together with 
the decaying mode. 
Alcubierre et al. \cite{potsdam9908} found that the 
adjustment 
due to the  momentum constraint 
is crucial for obtaining stability.  We 
think  that they picked up this error growing mode. 
Fortunately at the BSSN 
limit ($\kappa=1$), this error growing mode disappears and turns into 
a propagation mode. 
\item No ${\cal H}$-adjustment case.  The set 
$\kappa_{adj}=( 1,1,1,\kappa,1,1,1,1) $ gives 
$$ 
CAFs=( 0 \, (\times 3), \pm \sqrt{-  k^2} \, \mbox{(3 pairs)} ), 
$$ 
independently to $\kappa$. 
Therefore the effect of ${\cal H}$-adjustment 
is unimportant according to this analysis, i.e. 
on flat space-time background.
(Remark again that we do not deny the effect of ${\cal H}$-adjustment
 in other situation.)
\end{enumerate} 
These tests are on the effects of adjustments. 
We will consider whether much 
better adjustments are possible in the next section.

We list the above results in Table \ref{table1}. 
(Table \ref{table1} includes a column of diagonalizability of 
constraint propagation matrix $M$, of which importance was pointed
out in \cite{diagCP}. )
The most characteristic points of the above are No. 5 and No.6 
that 
denote the contribution of the momentum constraint adjustment and 
the importance of the new variable $\tilde{\Gamma}^i$. 
It is quite interesting that the unadjusted BSSN equations (case 1)
does not have apparent advantages from the ADM system.
As we showed in the case 5 and 6, if we missed a particular
adjustment,  then the expected stability behavior occationally 
gets worse
than the starting ADM system.  
Therefore we conclude that the better stability of the 
BSSN formulation is obtained by their adjustments in the equations, 
and the combination of the adjustments is in a good balance. 
That is, a lack of their adjustments might fail to obtain the stability
of their system.


\section{Proposals of Improved BSSN systems} \label{sec4} 
\label{sec4}
In this section, we consider the possibility whether we can obtain a 
system which has much better properties; 
whether more pure imaginary CAFs 
or negative real CAFs. 

\subsection{Heuristic examples} 

(A) A system which has 8 pure imaginary CAFs: \\ 
One direction is to seek a set of equations which 
make fewer zero CAFs than the standard BSSN case
(No.2 in the previous section). 
Using the same set of adjustments in 
(\ref{eqmMinkow1})-(\ref{eqmMinkow5}), 
CAFs are written in general 
\begin{eqnarray*} 
CAFs &=& \Big(0,\pm\sqrt{-k^2 \kappa_{A1} 
\kappa_{\tilde{\Gamma} 2}} \, \mbox{(2 pairs)}, 
\\&& 
\pm \mbox{complicated expression}, 
\\&& 
\pm \mbox{complicated expression} \Big). 
\end{eqnarray*} 
The terms in the first line certainly give 
four pure imaginary CAFs (two 
positive and negative real pairs) if 
$\kappa_{A1}\kappa_{\tilde{\Gamma} 2}>0 \, (<0)$. 
Keeping this in  mind, by choosing 
$\kappa_{adj}=(1,1,1, 1,1,\kappa,1,1)$, we find 
\begin{eqnarray*} 
CAFs&=&\Big(0,\pm\sqrt{-k^2 } \, \mbox{(2 pairs)}, 
\\&& 
\pm\sqrt{-k^2 (2+\kappa+|\kappa-4|)/6}, 
\\&& 
\pm\sqrt{-k^2 (2+\kappa-|\kappa-4|)/6}, 
\Big). 
\end{eqnarray*} 
Therefore the adjustment 
$\kappa_{adj}=( 1,1,1, 1,1,4,1,1 )$ gives 
$$ 
CAFs=\Big(0,\pm\sqrt{-k^2 } \, \mbox{(4 pairs)} \Big), 
$$ 
which is one step advanced from BSSN's according our guidelines. 

We note that 
such a system can be obtained in many ways, e.g. 
$\kappa_{adj}=(0,0,1,0,2,1,0,1/2)$ also gives four pairs of pure 
imaginary CAFs. 
\\ 

(B) A system which has negative real CAF: \\ 
One criterion to obtain a decaying constraint mode 
(i.e. an asymptotically 
constrained system) is to adjust an evolution equation as it breaks 
time reversal symmetry \cite{adjADM,adjADMsch}. 
For example, we consider an additional adjustment 
to the BSSN equation as 
\begin{equation} 
\partial_t \tilde{\gamma}_{ij} = \partial_t^B 
\tilde{\gamma}_{ij}+\kappa_{SD} 
\alpha \tilde{\gamma}_{ij} {\cal H}, 
\end{equation} 
which is a similar adjustment of the simplified 
Detweiler-type \cite{detweiler} 
that was discussed in \cite{adjADMsch}. 
The first order constraint propagation 
equations on the flat background space-time become 
\begin{eqnarray*} 
\partial_t  {}^{\!(1)\!\!} {\cal H}&=& 
\partial_j\partial_j {}^{\!(1)\!\!} {\cal A} 
-(3/2)\kappa_{SD}\partial_j\partial_j  {}^{\!(1)\!\!} 
{\cal H}, 
\\ 
\partial_t {}^{\!(1)\!\!} {\cal M}_i 
&=& 
(1/6)\partial_i  {}^{\!(1)\!\!} {\cal H} 
+(1/2)\partial_j\partial_j  {}^{\!(1)\!\!} {\cal G}^i, 
\\ 
\partial_t {}^{\!(1)\!\!} {\cal G}^i  &=& 
-\partial_i {}^{\!(1)\!\!} {\cal A} 
+(1/2)\kappa_{SD}\partial_i {}^{\!(1)\!\!} {\cal H} 
+2 {}^{\!(1)\!\!} {\cal M}_i, 
\\ 
\partial_t  {}^{\!(1)\!\!} {\cal A} &=& 
- (\partial_j\partial_j{}^{\!(1)\!\!} \alpha)^{TF} 
+ ({}^{\!(1)\!\!} R^{BSSN}_{jj})^{TF}, 
\\ 
\partial_t  {}^{\!(1)\!\!} {\cal S} 
&=& 
-2 {}^{\!(1)\!\!} {\cal A}+3\kappa_{SD} 
{}^{\!(1)\!\!} {\cal H}, 
\end{eqnarray*} 
where we wrote only additional terms to 
(\ref{CPMinkow1})-(\ref{CPMinkow5}). 
The CAFs become 
\begin{eqnarray*} 
CAFs &=&(0 \, (\times2), 
\\&& 
\pm\sqrt{-k^2} \mbox{(3 pairs)}, \, (3/2)k^2 \kappa_{SD}), 
\end{eqnarray*} 
in which the last one becomes negative real if 
$\kappa_{SD} <0$.  \\ 

(C) Combination of above (A) and (B) \\ 
Naturally we next consider both  adjustments: 
\begin{eqnarray} 
\partial_t \tilde{\gamma}_{ij} &=& \partial_t^B 
\tilde{\gamma}_{ij}+\kappa_{SD} 
\alpha \tilde{\gamma}_{ij} {\cal H} 
\\ 
\partial_t \tilde{A}_{ij} 
&=& 
\partial_t^B \tilde{A}_{ij} 
-\kappa_{8}\alpha e^{-4\varphi}\tilde{\gamma}_{ij} \partial_k{\cal G}^k 
\end{eqnarray} 
where the second one produces the 8 pure imaginary CAFs. 
The additional terms in the constraint propagation equations 
(\ref{CPMinkow1})-(\ref{CPMinkow5}) are 
\begin{eqnarray*} 
\partial_t {}^{\!(1)\!\!}  {\cal H}&=& 
\partial_j\partial_j{}^{\!(1)\!\!}  {\cal A} 
-(3/2)\kappa_{SD}\partial_j\partial_j 
{}^{\!(1)\!\!} {\cal H}, 
\\ 
\partial_t  {}^{\!(1)\!\!} {\cal M}_i&=& 
(1/6)\partial_i  {}^{\!(1)\!\!} {\cal H} 
+(1/2)\partial_j\partial_j {}^{\!(1)\!\!}  {\cal G}^i 
\nonumber \\&& 
-\kappa_{8}\partial_i\partial_k {}^{\!(1)\!\!} {\cal G}^k, 
\\ 
\partial_t  {}^{\!(1)\!\!} {\cal G}^i &=& 
-\partial_i {}^{\!(1)\!\!} {\cal A} 
+(1/2)\kappa_{SD}\partial_i {}^{\!(1)\!\!} {\cal H} 
+2 {}^{\!(1)\!\!} {\cal M}_i, 
\\ 
\partial_t  {}^{\!(1)\!\!} {\cal A} &=& 
-3\kappa_{8}\partial_k{}^{\!(1)\!\!}  {\cal G}^k,  
\\ 
\partial_t  {}^{\!(1)\!\!} {\cal S} 
&=& 
-2{}^{\!(1)\!\!}{\cal A}+3\kappa_{SD}{}^{\!(1)\!\!}{\cal H}.
\end{eqnarray*} 
We then obtain 
\begin{eqnarray*} 
CAFs&=&\Big( 
0,\pm\sqrt{-k^2} \, \mbox{(3 pairs)}, 
\\&& 
(3/4)k^2 \kappa_{SD} 
\pm \sqrt{k^2(-\kappa_{8}+(9/16) k^2 \kappa^2_{SD})} 
\Big) 
\end{eqnarray*} 
which reproduces case (A) when $\kappa_{SD}=0,\kappa_{8}=1$, 
and case (B) when $\kappa_{8}=0$. 
These CAFs can become 
(0, pure imaginary (3 pairs), complex numbers 
with a negative real part (1 pair)), 
with an appropriate combination of $\kappa_{8}$ and $\kappa_{SD}$. 

\begin{widetext} 
\subsection{Possible adjustments} 
In order to break time reversal symmetry of the evolution equations 
\cite{adjADM,adjADMsch,BFHR}, 
the possible simple adjustments are 
(1) to add ${\cal H}$, ${\cal S}$ or ${\cal G}^i$ terms 
to the equations of 
$\partial_t \phi$, $\partial_t \tilde{\gamma}_{ij}$, or 
$\partial_t \tilde{\Gamma}^i$, or (2) to add ${\cal M}_{i}$ 
or ${\cal A}$ terms to 
$\partial_t  K$ or $\partial_t {\tilde A}_{ij}$. 
We write them generally, including the above proposal (B),  as 
\begin{eqnarray} 
\partial_t \phi &=& \partial_t^B \phi 
+ \kappa_{\phi{\cal H}} \, \alpha{\cal H} 
+ \kappa_{\phi{\cal G}}\,  \alpha{\tilde D}_k{\cal G}^k,  
\label{negadj1} 
\\ 
\partial_t \tilde{\gamma}_{ij} &=& \partial_t^B \tilde{\gamma}_{ij} 
+ \kappa_{SD} \, \alpha\tilde{\gamma}_{ij}{\cal H} 
+ \kappa_{\tilde{\gamma}{\cal G} 1} \, 
\alpha\tilde{\gamma}_{ij}{\tilde D}_k{\cal G}^k 
+ \kappa_{\tilde{\gamma}{\cal G} 2} \, 
\alpha\tilde{\gamma}_{k(i}{\tilde D}_{j)}{\cal G}^k 
+ \kappa_{\tilde{\gamma} {\cal S} 1}\, 
\alpha\tilde{\gamma}_{ij}{\cal S} 
+ \kappa_{\tilde{\gamma}{\cal S} 2} \, 
\alpha{\tilde D}_i{\tilde D}_j{\cal S}, 
\\ 
\partial_t K &=&\partial_t^B K 
+ \kappa_{K {\cal M}} \, 
\alpha \tilde{\gamma}^{jk}({\tilde D}_j{\cal M}_k), 
\\ 
\partial_t {\tilde A}_{ij}&=&\partial_t^B {\tilde A}_{ij} 
+ \kappa_{A {\cal M} 1} \, 
\alpha \tilde{\gamma}_{ij} ({\tilde D}^k{\cal M}_k) 
+ \kappa_{A {\cal M} 2} \, 
\alpha ({\tilde D}_{(i}{\cal M}_{j)}) 
+ \kappa_{A {\cal A}1} \, 
\alpha \tilde{\gamma}_{ij} {\cal A} 
+ \kappa_{A {\cal A}2} \, 
\alpha {\tilde D}_i{\tilde D}_j{\cal A}, 
\\ 
\partial_t \tilde{\Gamma}^i&=&\partial_t^B \tilde{\Gamma}^i 
+ \kappa_{\tilde{\Gamma} {\cal H}} \, 
\alpha {\tilde D}^i{\cal H} 
+ \kappa_{\tilde{\Gamma} {\cal G} 1} \, 
\alpha {\cal G}^i 
+ \kappa_{\tilde{\Gamma} {\cal G} 2} \, 
\alpha {\tilde D}^j{\tilde D}_j{\cal G}^i 
+ \kappa_{\tilde{\Gamma} {\cal G} 3} \, 
\alpha {\tilde D}^i{\tilde D}_j{\cal G}^j, 
\label{negadj5} 
\end{eqnarray} 
where $\kappa$s are possible multipliers  (all $\kappa=0$ reduce the 
system the standard BSSN evolution equations). 

\end{widetext} 

We show the effects of each terms in Table \ref{table2}. 
The CAFs in the table are on the 
flat space background.  We see several terms produce 
negative real-part in CAFs, which might improve the stability than 
the previous system.  
(Table \ref{table2} includes again a column of diagonalizability of 
constraint propagation matrix $M$. Diagonalizable ones are expected to 
reflect the predictions from eigenvalue analysis. That is, the eigenvalue
analysis with diagonalizable ones definitely
avoid the diverging possibility in constraint propagation when it 
includes degenerated CAFs.  See \cite{diagCP}). 
For the readers convenience, we list 
up several best candidates here. 

(D) { A system which has 7 negative CAFs}\\ 
Simply adding $\tilde{D}_{(i}{\cal M}_{j)}$ term to 
$\partial_t \tilde{A}_{ij}$ equation, say 
\begin{equation} 
\partial_t \tilde{A}_{ij}=\partial^{BSSN}_t \tilde{A}_{ij} 
+\kappa_{A {\cal M} 2} \alpha (\tilde{D}_{(i}{\cal M}_{j)}) 
\end{equation} 
with 
$\kappa_{A {\cal M} 2}>0$, CAFs on the flat background are 7 negative 
real CAFs. \\ 

(E) { A system which has 6 negative CAFs}\\ 
The below two adjustments will make 
6 negative real CAFs, while they also produce one positive real CAF 
(a constraint violating mode).  The effectiveness is not clear 
at this moment, but we think they are worth to be tested in numerical 
experiments. 

(E1) With $\kappa_{\tilde{\gamma} {\cal G} 2} <0$, 
\begin{equation} 
\partial_t \tilde{\gamma}_{ij} =\partial^{BSSN}_t \tilde{\gamma}_{ij} 
+ \kappa_{\tilde{\gamma}{\cal G} 2} \, 
\alpha\tilde{\gamma}_{k(i}{\tilde D}_{j)}{\cal G}^k.  
\end{equation}

(E2) With $\kappa_{\tilde{\Gamma} {\cal G} 2} <0$. 
\begin{equation} 
\partial_t \tilde{\Gamma}^i = \partial^{BSSN}_t \tilde{\Gamma}^i 
+ \kappa_{\tilde{\Gamma} {\cal G} 2} \, 
\alpha {\tilde D}^j{\tilde D}_j{\cal G}^i. 
\end{equation}

\section{Concluding remarks} \label{Summary} 
Applying the constraint propagation analysis, 
we tried to understand why and how the so-called BSSN 
(Baumgarte-Shapiro-Shibata-Nakamura) 
re-formulation works better than the standard ADM 
equations in general relativistic numerical simulations. 
Our strategy was to evaluate eigenvalues of 
the constraint propagation equations in their Fourier modes, 
which method succeeded to explain 
the stability properties in many other 
systems in our series of works. 

We have studied step-by-step where the replacements in the equations 
affect and/or newly added constraints work, by checking whether the 
error of constraints (if it exists) will decay or propagate away. 
Alcubierre et al \cite{potsdam9908} pointed out  
the importance of the 
replacement (adjustment) to the evolution equation 
using the momentum constraint, 
and our analysis clearly explains why they concluded 
this is the key. 
Not only this adjustment, we found, 
but also other adjustments and other 
introductions of new constraints also contribute 
to making the evolution system more stable. 
We found that if we missed a particular adjustment, 
then the expected stability behavior occationally gets worse than the 
ADM system. 
We further propose other adjustments of 
the set of equations which may have better features 
for numerical treatments. 

The discussion in this article was only in the flat 
background space-time, and may not be applicable directly 
to the general numerical simulations.  However, 
we 
rather 
believe that the general fundamental aspects of constraint
propagation analysis are already 
revealed in this 
article.  This is because, 
for the ADM and its adjusted cases, we found that 
the better formulations 
in the flat background are also better 
in the Schwarzschild space-time, while 
there are differences on the effective adjusting multipliers or the 
effective coordinate ranges\cite{adjADMsch,adjADM}. 

We have not shown any numerical tests here.  However, 
recently, the proposal (B) in \S \ref{sec4} was examined 
numerically using linear wave initial data 
and confirmed to be effective for controlling the violation 
of the Hamiltonian constraint with our predicted multiplier signature 
\cite{masarushibata}. 
The systematic numerical comparisons between different formulations 
are underway 
\cite{mexico}, and we expect to have a chance 
to report them in near future.  We are also trying to explain 
the stability of Laguna-Shoemaker's implimented BSSN system 
\cite{PabloDeirdre} using 
the constraint propagation analysis.

There may not be the almighty formulation for any models in numerical 
relativity, but we believe our guidelines to find a 
better formulation in a systematic way will contribute 
a progress of this field. 
We hope the predictions in this paper 
will help the community to make further improvements.

\section*{Acknowledgements} 
HS thanks T. Nakamura and M. Shibata for their comments, and he 
appreciates the hospitality 
of the Centre for Gravitational Wave Physics, the Pennsylvania State 
University, where the 
part of this work has been done. 
HS is supported by the special postdoctoral researchers' program at 
RIKEN. 
This work was supported partially by the Grant-in-Aid for Scientific 
Research Fund of Japan Society of the Promotion of Science, No. 14740179. 



\begin{widetext} 

\appendix 
\section{Full set of BSSN constraint propagation equations} 
The  constraint propagation equations of the BSSN system can be written 
as follows. 
\begin{eqnarray} 
\lefteqn{ \partial_t{\cal H}= 
\Big( 
(2/3)\alpha K 
+(2/3)\alpha {\cal A} 
+\beta^k\partial_k 
\Big){\cal H} 
+\Big( 
-4e^{-4\varphi}\alpha (\partial_k\varphi)\tilde\gamma^{kj} 
-2e^{-4\varphi}(\partial_k\alpha) \tilde\gamma^{jk} 
\Big){\cal M}_j 
} 
\nonumber \\&& 
+\Big( 
-2\alpha e^{-4\varphi} \tilde{A}^k{}_j\partial_k 
-\alpha e^{-4\varphi} (\partial_j\tilde{A}_{kl})\tilde\gamma^{kl} 
-e^{-4\varphi}(\partial_j\alpha){\cal A} 
-e^{-4\varphi}\beta^k\partial_k\partial_j 
\nonumber \\&& 
-(1/2)e^{-4\varphi}\beta^k \tilde\gamma^{-1} 
(\partial_j{\cal S})\partial_k 
+(1/6)e^{-4\varphi}\tilde\gamma^{-1} 
(\partial_j\beta^k)(\partial_k{\cal S}) 
-(2/3)e^{-4\varphi}(\partial_k\beta^k)\partial_j 
\Big){\cal G}^j 
\nonumber \\&& 
+\Big( 
2 \alpha e^{-4\varphi}\tilde\gamma^{-1}\tilde\gamma^{lk} 
(\partial_l\varphi){\cal A} 
\partial_k +(1/2)\alpha e^{-4\varphi}\tilde\gamma^{-1} 
(\partial_l{\cal A})\tilde\gamma^{lk}\partial_k 
+(1/2)e^{-4\varphi}\tilde\gamma^{-1} 
(\partial_l\alpha)\tilde\gamma^{lk}{\cal A}\partial_k 
\nonumber \\&& 
+(1/2)e^{-4\varphi}\tilde\gamma^{-1}\beta^m\tilde\gamma^{lk} 
\partial_m\partial_l\partial_k 
-(5/4)e^{-4\varphi}\tilde\gamma^{-2}\beta^m\tilde\gamma^{lk} 
(\partial_m{\cal S})\partial_l\partial_k + 
e^{-4\varphi}\tilde\gamma^{-1}\beta^m 
(\partial_m\tilde\gamma^{lk})\partial_l\partial_k 
\nonumber \\&& 
+(1/2)e^{-4\varphi}\tilde\gamma^{-1}\beta^i 
(\partial_j\partial_i\tilde\gamma^{jk})\partial_k 
+(3/4)e^{-4\varphi}\tilde\gamma^{-3}\beta^i\tilde\gamma^{jk} 
(\partial_i{\cal S})(\partial_j{\cal S})\partial_k 
-(3/4)e^{-4\varphi}\tilde\gamma^{-2}\beta^i 
(\partial_i\tilde\gamma^{jk})(\partial_j{\cal S})\partial_k 
\nonumber \\&& 
+(1/3)e^{-4\varphi}\tilde\gamma^{-1}\tilde\gamma^{pj} 
(\partial_j\beta^k)\partial_p\partial_k 
\nonumber \\&& 
-(5/12)e^{-4\varphi}\tilde\gamma^{-2}\tilde\gamma^{jk} 
(\partial_k\beta^i)(\partial_i{\cal S})\partial_j 
+(1/3)e^{-4\varphi}\tilde\gamma^{-1} 
(\partial_k\tilde\gamma^{ij})(\partial_j\beta^k)\partial_i 
-(1/6)e^{-4\varphi}\tilde\gamma^{-1}\tilde\gamma^{mk} 
(\partial_k\partial_l\beta^l)\partial_m \Big){\cal S} 
\nonumber \\&& 
+\Big( 
(4/9)\alpha K {\cal A} 
-(8/9)\alpha K^2 
+(4/3)\alpha e^{-4\varphi}(\partial_i\partial_j\varphi)\tilde\gamma^{ij} 
+(8/3)\alpha e^{-4\varphi} 
(\partial_k\varphi)(\partial_l\tilde\gamma^{lk}) 
\nonumber \\&& 
+\alpha e^{-4\varphi}(\partial_j\tilde\gamma^{jk})\partial_k 
+8 \alpha e^{-4\varphi}\tilde\gamma^{jk}(\partial_j\varphi)\partial_k 
+\alpha e^{-4\varphi}\tilde\gamma^{jk}\partial_j\partial_k 
+8e^{-4\varphi}(\partial_l\alpha)(\partial_k\varphi)\tilde\gamma^{lk} 
+e^{-4\varphi}(\partial_l\alpha)(\partial_k\tilde\gamma^{lk}) 
\nonumber \\&& 
+2e^{-4\varphi}(\partial_l\alpha)\tilde\gamma^{lk}\partial_k 
+e^{-4\varphi}\tilde\gamma^{lk}(\partial_l\partial_k\alpha) 
\Big){\cal A}, 
\\ 
\lefteqn{ 
\partial_t{\cal M}_i= 
\Big( 
-(1/3)(\partial_i\alpha) 
+(1/6)\partial_i 
\Big){\cal H} 
+\alpha K {\cal M}_i 
+\Big( 
\alpha e^{-4\varphi}\tilde\gamma^{km}(\partial_k \varphi) 
(\partial_j\tilde\gamma_{mi})  -(1/2)\alpha 
e^{-4\varphi}\tilde\Gamma^m_{kl}\tilde\gamma^{kl} 
(\partial_j\tilde\gamma_{mi})} 
\nonumber \\&& 
+(1/2)\alpha 
e^{-4\varphi}\tilde\gamma^{mk}(\partial_k\partial_j\tilde\gamma_{mi}) 
+(1/2)\alpha e^{-4\varphi}\tilde\gamma^{-2}(\partial_i{\cal 
S})(\partial_j{\cal S})  -(1/4)\alpha 
e^{-4\varphi}(\partial_i\tilde\gamma_{kl}) 
(\partial_j\tilde\gamma^{kl}) +\alpha 
e^{-4\varphi}\tilde\gamma^{km} 
(\partial_k \varphi) \tilde\gamma_{ji}\partial_m 
\nonumber \\&& 
+\alpha e^{-4\varphi} (\partial_j \varphi)\partial_i 
-(1/2)\alpha 
e^{-4\varphi}\tilde\Gamma^m_{kl}\tilde\gamma^{kl} 
\tilde\gamma_{ji}\partial_m 
+\alpha e^{-4\varphi}\tilde\gamma^{mk} 
\tilde\Gamma_{ijk}\partial_m +(1/2)\alpha 
e^{-4\varphi}\tilde\gamma^{lk}\tilde\gamma_{ji}\partial_k\partial_l 
\nonumber \\&& 
+(1/2)e^{-4\varphi}\tilde\gamma^{mk}(\partial_j\tilde\gamma_{im}) 
(\partial_k\alpha) +(1/2)e^{-4\varphi} (\partial_j\alpha)\partial_i 
+(1/2)e^{-4\varphi} 
\tilde\gamma^{mk}\tilde\gamma_{ji}(\partial_k\alpha)\partial_m 
\Big){\cal G}^j 
\nonumber \\&& 
+\Big( 
-\tilde{A}^k{}_i(\partial_k\alpha) 
+(1/9)(\partial_i\alpha)K 
+(4/9)\alpha (\partial_i K) 
+(1/9)\alpha K \partial_i 
-\alpha \tilde{A}^k{}_i \partial_k 
\Big){\cal A}, 
\\ 
\lefteqn{ 
\partial_t{\cal G}^i 
= 
2\alpha \tilde\gamma^{ij}{\cal M}_j 
+\Big( 
-(1/2)\beta^k\tilde\gamma^{il}\tilde\gamma^{-2}(\partial_l{\cal 
S})\partial_k 
-(1/2)\beta^k\tilde\gamma^{in} 
(\partial_k\tilde\gamma_{mn})\tilde\gamma^{ml} 
\tilde\gamma^{-1}\partial_l 
+(1/2)\beta^k\tilde\gamma^{il}\tilde\gamma^{-1}\partial_l\partial_k } 
\nonumber \\&& 
-(1/2)(\partial_m\beta^i)\tilde\gamma^{mk}\tilde\gamma^{-1}\partial_k 
+(1/3)(\partial_l\beta^l)\tilde\gamma^{ik}\tilde\gamma^{-1}\partial_k 
\Big){\cal S} 
+\Big( 
+4\alpha \tilde\gamma^{ij}(\tilde{D}_j \varphi) 
-\alpha\tilde\gamma^{ij}\partial_j 
-(\partial_k\alpha)\tilde\gamma^{ik} 
\Big){\cal A}, 
\\ 
\lefteqn{ 
\partial_t{\cal S}= 
+\beta^k(\partial_k{\cal S}) 
-2\alpha\tilde\gamma{\cal A},  
} 
\\ 
\lefteqn{\partial_t {\cal A}= 
\Big(\alpha K 
+\beta^k\partial_k 
\Big){\cal A}. 
}
\end{eqnarray} 
The flat background linear order equations, 
(\ref{CPMinkow1})-(\ref{CPMinkow5}), were obtained from these 
expression.


\begin{table}[htb] 
\begin{tabular}{ll|ccccc||c|l} 
\hline 
No.  & & 
\multicolumn{5}{l||}{Constraints (number of components)}  
& diag? & CAFs
\\ 
in text.  & & ${\cal H}$ (1) &  ${\cal M}_i$ (3) 
&  ${\cal G}^i$ (3) & 
${\cal A} $ (1) &  ${\cal S} 
$ (1) & &
in Minkowskii background 
\\ 
\hline  &&&&&&&& \\ 
0.& standard ADM & use & use & - & -& -& yes & $(0,0,\Im,\Im)$ 
\\ 
1.& BSSN no adjustment & use & use & use & use & use & yes & 
$(0,0,0,0,0,0,0,\Im,\Im)$ 
\\ 
2. &the BSSN   & use+adj & use+adj & use+adj & use+adj & use+adj & no & 
$(0,0,0,\Im,\Im,\Im,\Im,\Im,\Im)$ 
\\ 
&&&&&&& \\ 
\hline 
&&&&&&& \\ 
3.& no ${\cal S} $ adjustment   & use+adj & use+adj 
& use+adj & use+adj & 
use  & no & 
no difference in flat background 
\\ 
4.& no ${\cal A} $ adjustment    & use+adj & use+adj 
& use+adj & use & use+adj & no 
& $(0,0,0,\Im,\Im,\Im,\Im,\Im,\Im)$ 
\\ 
5. & no ${\cal G}^i $ adjustment    & use+adj & use+adj 
& use & use+adj & 
use+adj & no & $(0,0,0,0,0,0,0,\Im,\Im)$ 
\\ 
6. &no ${\cal M}_i $ adjustment   & use+adj & use  & use+adj 
& use+adj & 
use+adj 
& no & $(0,0,0,0,0,0,0,\Re,\Re)$ 
\\ 
7. &no ${\cal H} $ adjustment   & use & use+adj  & use+adj 
& use+adj & 
use+adj & no
& $(0,0,0,\Im,\Im,\Im,\Im,\Im,\Im)$ 
\\ 
&&&&&&& \\ 
\hline 
\end{tabular} 
\caption{ 
Summary of \S \ref{sec3c}: contributions of adjustments terms and 
effects of introductions of new constraints 
in the BSSN system. 
The center column indicates whether each constraints are taken as 
a component of constraints in each constraint propagation analysis 
(`use'), and whether each adjustments are on (`adj'). 
The column `diag?' indicates
diagonalizability of the constraint propagation matrix. 
The right column shows CAFs, 
where 
$\Im$ and $\Re$ means pure imaginary and real eigenvalue, 
respectively. 
No.0 (standard ADM) is shown in \cite{adjADM}. 
}\label{table1} 
\end{table} 
\begin{table}[h] 
\begin{tabular}{ll|l|c|ll} 
\hline 
\multicolumn{2}{c|}{adjustment} & 
\multicolumn{1}{c|}{CAFs} & diag? & 
\multicolumn{2}{|c}{effect of the adjustment} 
\\ 
\hline 
$\partial_t \phi$ & 
$\kappa_{\phi{\cal H}} \, \alpha {\cal H}$ & 
$(0,0,\pm\sqrt{-k^2}(*3),8\kappa_{\phi{\cal H}}k^2)$ & no & 
$\kappa_{\phi{\cal H}}<0$ makes 1 Neg. & 
\\ 
$\partial_t \phi$ & 
$\kappa_{\phi{\cal G}} \, \alpha{\tilde D}_k{\cal G}^k$& 
$(0,0,\pm\sqrt{-k^2}(*2)$, long expressions) & yes &
$\kappa_{\phi{\cal G}}<0$ makes 2 Neg. 1 Pos. & 
\\ 
$\partial_t \tilde\gamma_{ij}$ & 
$\kappa_{SD} \, \alpha\tilde{\gamma}_{ij}{\cal H} $& 
$(0,0,\pm\sqrt{-k^2}(*3),(3/2)\kappa_{SD}k^2)$ & yes &
$\kappa_{SD}<0$ makes 1 Neg. & Case (B) 
\\ 
$\partial_t \tilde\gamma_{ij}$ & 
$\kappa_{\tilde{\gamma}{\cal G} 1} \, 
\alpha\tilde{\gamma}_{ij}{\tilde D}_k{\cal G}^k$& 
$(0,0,\pm\sqrt{-k^2}(*2)$, long expressions)& yes &
$\kappa_{\tilde{\gamma}{\cal G} 1}>0$ makes 1 Neg. 
\\ 
$\partial_t \tilde\gamma_{ij}$ & 
$\kappa_{\tilde{\gamma}{\cal G} 2} \, 
\alpha\tilde{\gamma}_{k(i}{\tilde D}_{j)}{\cal G}^k$& 
\begin{tabular}{l} 
(0,0, {\footnotesize
$(1/4) k^2 \kappa_{\tilde{\gamma}{\cal G} 2} \pm \sqrt{k^2(-1+k^2
\kappa_{\tilde{\gamma}{\cal G} 2}/16)}(*2)$}, \\ 
long expressions)
\end{tabular}& yes &
$\kappa_{\tilde{\gamma}{\cal G} 2}<0$ makes 6 Neg. 1 Pos. & Case (E1) 
\\ 
$\partial_t \tilde\gamma_{ij}$ & 
$\kappa_{\tilde{\gamma} {\cal S} 1} \, 
\alpha\tilde{\gamma}_{ij}{\cal S}$& 
$(0, 0,\pm\sqrt{-k^2}(*3),3\kappa_{\tilde{\gamma}{\cal S} 1})$& no &
$\kappa_{\tilde{\gamma}{\cal S} 1}<0$ makes 1 Neg.   & 
\\ 
$\partial_t \tilde\gamma_{ij}$ & 
$\kappa_{\tilde{\gamma}{\cal S} 2} \, 
\alpha{\tilde D}_i{\tilde D}_j{\cal S}$& 
$(0, 0,\pm\sqrt{-k^2}(*3),-\kappa_{\tilde{\gamma}{\cal S} 2} k^2) $& no &
$\kappa_{\tilde{\gamma}{\cal S} 2}>0$ makes 1 Neg.  & 
\\ 
$\partial_t K$ & 
$\kappa_{K {\cal M}} \,  \alpha 
\tilde{\gamma}^{jk}({\tilde D}_j{\cal M}_k)$& 
\begin{tabular}{l} 
$(0,0,0,\pm\sqrt{-k^2}(*2), $ \\ 
~~$ (1/3) \kappa_{K {\cal M}} k^2\pm(1/3) 
\sqrt{k^2(-9+k^2\kappa_{K {\cal M}}^2)})$ 
\end{tabular} 
& no & $\kappa_{K {\cal M}}<0$ makes 2 Neg. & 
\\ 
$\partial_t \tilde{A}_{ij}$ & 
$\kappa_{A {\cal M} 1} \, 
\alpha \tilde{\gamma}_{ij} ({\tilde D}^k{\cal M}_k)$& 
$(0,0,\pm\sqrt{-k^2}(*3),-\kappa_{A {\cal M} 1}k^2)$ & yes &
$\kappa_{A {\cal M} 1}>0$ makes 1 Neg. & 
\\ 
$\partial_t \tilde{A}_{ij}$ & 
$ \kappa_{A {\cal M} 2} \,  \alpha ({\tilde D}_{(i}{\cal M}_{j)})$& 
\begin{tabular}{l} 
(0,0, {\footnotesize
$-k^2 \kappa_{A {\cal M} 2}/4 \pm \sqrt{k^2(-1+k^2\kappa_{A {\cal M} 2}/16)}(*2)$
, }\\
 long expressions)
\end{tabular}& yes &
$\kappa_{A {\cal M} 2}>0$ makes 7 Neg  & Case (D) 
\\ 
$\partial_t \tilde{A}_{ij}$ & 
$\kappa_{A {\cal A}1} \,  \alpha \tilde{\gamma}_{ij} {\cal A}$& 
$(0,0,\pm\sqrt{-k^2}(*3),3\kappa_{A {\cal A}1})$ & yes &
$\kappa_{A {\cal A}1}<0$ makes 1 Neg. & 
\\ 
$\partial_t \tilde{A}_{ij}$ & 
$\kappa_{A {\cal A}2} \,  \alpha {\tilde D}_i{\tilde D}_j{\cal A}$& 
$(0,0,\pm\sqrt{-k^2}(*3),-\kappa_{A {\cal A}2}k^2)$ & yes &
$\kappa_{A {\cal A}2}>0$ makes 1 Neg. & 
\\ 
$\partial_t \tilde{\Gamma}^i$ & 
$\kappa_{\tilde{\Gamma} {\cal H}} \,  \alpha {\tilde D}^i{\cal H}$& 
$(0,0,\pm\sqrt{-k^2}(*3),-\kappa_{A {\cal A}2}k^2)$ & no &
$\kappa_{\tilde{\Gamma} {\cal H}} >0$ makes 1 Neg. & 
\\ 
$\partial_t \tilde{\Gamma}^i$ & 
$ \kappa_{\tilde{\Gamma} {\cal G} 1} \,  \alpha {\cal G}^i$& 
$(0,0,(1/2)\kappa_{\tilde{\Gamma} {\cal G} 1}\pm\sqrt{-k^2+\kappa_{\tilde{\Gamma}
{\cal G} 1}^2}(*2)$ , long exp.)& yes &
$\kappa_{\tilde{\Gamma} {\cal G} 1}<0$ makes 6 Neg. 1 Pos. & Case (E2) 
\\ 
$\partial_t \tilde{\Gamma}^i$ & 
$ \kappa_{\tilde{\Gamma} {\cal G} 2} \, 
\alpha {\tilde D}^j{\tilde D}_j{\cal G}^i$& 
$(0,0,-(1/2)\kappa_{\tilde{\Gamma} {\cal G} 2}\pm\sqrt{-k^2+\kappa_{\tilde{\Gamma}
{\cal G} 2}^2}(*2)$ , long exp.)& yes &
$\kappa_{\tilde{\Gamma} {\cal G} 2}>0$ makes 2 Neg. 1 Pos. & 
\\ 
$\partial_t \tilde{\Gamma}^i$ & 
$ \kappa_{\tilde{\Gamma} {\cal G} 3} \, 
\alpha {\tilde D}^i{\tilde D}_j{\cal G}^j$& 
$(0,0,-(1/2)\kappa_{\tilde{\Gamma} {\cal G} 3}\pm\sqrt{-k^2+\kappa_{\tilde{\Gamma}
{\cal G} 3}^2}(*2)$ , long exp.)& yes &
$\kappa_{\tilde{\Gamma} {\cal G} 3}>0$ makes 2 Neg. 1 Pos. & 
\\ 
\hline 
\end{tabular} 
\caption{ 
Possible adjustements which make a real-part 
CAFs negative  (\S \ref{sec4} B). 
The column of adjustments are nonzero multipliers in terms of 
(\ref{negadj1})-(\ref{negadj5}), which all violate 
time reversal symmetry of the equation. 
The column `diag?' indicates
diagonalizability of the constraint propagation matrix. 
Neg./Pos. means negative/positive respectively. 
}\label{table2} 
\end{table} 


\end{widetext}

%
\end{document}